\documentclass[12pt]{article}
\usepackage[cp1251]{inputenc}
\usepackage[T2A]{fontenc}
\usepackage[english]{babel}
\usepackage{amssymb}
\usepackage{amsmath}
\usepackage{amsfonts,mathrsfs,amscd}

\begin{document}
\title{On eigenvalues of the Landau Hamiltonian with a periodic electric potential}

\author{L.I.~Danilov}
\maketitle
\thispagestyle{empty}
\vspace{3mm}
\noindent{\it{\small Udmurt Federal Research Center of Ural Branch of the Russian Academy of Sciences, Izhevsk, Russia}}

\noindent{\it{\small danilov@udman.ru}}

\begin{abstract}
\noindent {\footnotesize  
We consider the Landau Hamiltonian $\widehat H_B+V$ on $L^2({\mathbb R}^2)$ with a periodic electric potential $V$. For every $m\in {\mathbb N}$ we prove that there exist nonconstant periodic electric potentials $V\in C^{\infty }({\mathbb R}^2;{\mathbb R})$ with zero mean values that analytically depend on a small parameter $\varepsilon \in {\mathbb R}$ such that the Landau level $(2m+1)B$ is an eigenvalue of the Hamiltonian (of infinite multiplicity) where $B>0$ is a strength of a homogeneous magnetic field.}\end{abstract}
\vskip 0.2cm
MSC 35P05

\section*{Introduction}

We consider the Landau Hamiltonian
$$
\widehat H_B+V=\left( -i\, \frac{\partial }{\partial x_1}\right) ^2+\left( -i\, \frac{\partial }{\partial x_2}-Bx_1\right) ^2+V
\eqno(0.1)
$$
acting on $L^2({\mathbb R}^2)$ where $V$ is a periodic electric potential $V\in L^2_{\mathrm {loc}}({\mathbb R}^2;{\mathbb R})$ and $B>0$ is the strength of a homogeneous magnetic field. The coordanates in ${\mathbb R}^2$ are defined with respect to some orthonomal basis $e_1,e_2$. For the zero electric potential $V\equiv 0$ the spectrum of the Landau Hamiltonian
$$
\widehat H_B=\left( -i\, \frac{\partial }{\partial x_1}\right) ^2+\left( -i\, \frac{\partial }{\partial x_2}-Bx_1\right) ^2
$$
consists of the eigenvalues $\lambda =(2m+1)B$, $m\in {\mathbb Z}_+\doteq {\mathbb N} \cup \{ 0\}$, of infinite multiplicity (Landau levels). Until recently, for a rational magnetic flux, it was not known whether eigenvalues can exist in the spectrum of operator (0.1) for nonconstant periodic electric potentials $V\in C^{\infty }({\mathbb R}^2;{\mathbb R})$ [1,2,3] (in [2] this problem was stated for nonconstant periodic potentials $V\in C({\mathbb R}^2;{\mathbb R})$). In [4], in the case of a rational magnetic flux, it was proved that there are nonconstant periodic electric potentials $V\in C^{\infty }({\mathbb R}^2;{\mathbb R})$ with zero mean values such that the operator (0.1) has an eigenvalue at the second Landau level $3B$. In this article for any $m\in {\mathbb N}$ we give examples of families of periodic electric potentials $V(x_1,x_2)=v(x_2)$ such that the operator $\widehat H_B+V$ has an eigenvalue at the $(m+1)$-th Landau level $(2m+1)B$; the nonconstant $T$-periodic functions $v(\cdot)$ with a given period $T>0$ and zero mean values analytically depend on a small parameter $\varepsilon \in {\mathbb R}$, the strength of a magnetic field $B$ also depends on the parameter $\varepsilon $. These examples imply that for any $m\in {\mathbb N}$ there are nonconstant periodic electric potentials $V\in C^{\infty }({\mathbb R}^2;{\mathbb R})$ with zero mean values such that the magnetic flux is rational and the operator $\widehat H_B+V$ has the eigenvalue $\lambda =(2m+1)B$.
\vskip 0.3cm

{\bf 0.1. Notation.} Let $E^1$ and $E^2$ be basis vectors of the non-degenerate lattice $\Lambda =\{ N_1E^1+N_2E^2: N_1,N_2\in {\mathbb Z}\} $; let $E^l_j=(E^l,e_j)$, $l,j=1,2$ (the inner product and the length of vectors from ${\mathbb R}^2$ are denoted by $|\cdot |$ and $(\cdot ,\cdot )$, respectively). By $K=\{ \xi _1E^1+\xi _2E^2:0\leqslant \xi _j\leqslant 1, j=1,2\} $ we denote the unit cell of the lattice $\Lambda $ of area $v(K)$ corresponding to the basis vectors $E^1,E^2$ where $v(\cdot )$ is the Lebesgue measure on ${\mathbb R}^2$, and let $\eta =(2\pi )^{-1}Bv(K)$ be the magnetic flux (defined for the unit cell $K$). We use the notation $\Lambda ^*$ for the reciprocal lattice in ${\mathbb R}^2$ with basis vectors $E^1_*, E^2_*\in {\mathbb R}^2$ for which $(E^{\, \mu },E^{\, \nu }_*)=\delta _{\mu \nu }$, ${\mu },{\nu }=1,2$, where $\delta _{\mu \nu }$ is the Kronecker delta.

For ${\mathcal B}={\mathbb R}$ or ${\mathcal B}={\mathbb C}$, let $L^p_{\Lambda }({\mathbb R}^2;{\mathcal B})$, $p\in [1,+\infty ]$, $C^n_{\Lambda }({\mathbb R}^2;{\mathcal B})$, $n\in {\mathbb Z}_+$, $C^{\infty }_{\Lambda }({\mathbb R}^2;{\mathcal B})$ and $H^s_{\Lambda }({\mathbb R}^2;{\mathcal B})$, $s\geqslant 0$, be the spaces of functions $W:{\mathbb R}^2\to {\mathcal B}$ from $L^p_{\mathrm {loc}}({\mathbb R}^2;{\mathcal B})$, $C^n({\mathbb R}^2;{\mathcal B})$, $C^{\infty }({\mathbb R}^2;{\mathcal B})$ and from Sobolev spaces $H^s_{\mathrm {loc}}({\mathbb R}^2;{\mathcal B})$, respectively, which are periodic with a period lattice $\Lambda $  ($C^0_{\Lambda }({\mathbb R}^2;{\mathcal B})=C_{\Lambda }({\mathbb R}^2;{\mathbb C})$ and $H^0_{\Lambda }({\mathbb R}^2;{\mathcal  B})=L^2_{\Lambda }({\mathbb R}^2;{\mathcal B})$). For functions $W\in L^p_{\Lambda }({\mathbb R}^2;{\mathbb C})$ and $W\in C_{\Lambda }({\mathbb R}^2;{\mathbb C})$ we define the norms $\| W\| _{L^p_{\Lambda }}\doteq \| W(\cdot \big|_K)\| _{L^p(K)}$ and $\| W\| _{C_{\Lambda }}\doteq \| W(\cdot \big|_K)\| _{C(K)}$.

Let $W_Y$ denote the Fourier coefficients of functions $W\in L^1_{\Lambda }({\mathbb R}^2;{\mathbb C})$:
$$
W_Y=(v(K))^{-1}\int_KW(x)\, e^{-i(Y,x)}\, dx,\quad Y\in 2\pi \Lambda ^*;
$$
$W_0$ is the mean value of the function $W$ (where $0\in 2\pi \Lambda ^*$ is the zero vector).

By $L^2_{\Lambda (T)}({\mathbb R};{\mathcal B})$,  $C_{\Lambda (T)}({\mathbb R};{\mathcal B})$, and $C^{\infty }_{\Lambda (T)}({\mathbb R};{\mathcal B})$, where $\Lambda (T)=T\mathbb Z$, $T>0$, and ${\mathcal B}={\mathbb R}$ or ${\mathcal B}={\mathbb C}$, we denote the spaces of $T$-periodic functions $w:{\mathbb R}\to {\mathcal B}$ from the spaces $L^2_{\mathrm {loc}}({\mathbb R};{\mathcal B})$, $C({\mathbb R};{\mathcal B})$, and $C^{\infty }({\mathbb R};{\mathcal B})$, respectively. For functions $w\in L^2_{\Lambda (T)}({\mathbb R};{\mathbb C})$ and $w\in C_{\Lambda (T)}({\mathbb R};{\mathbb C})$ we write  $\| w\| _{L^2_{\Lambda (T)}}\doteq \| w(\cdot \big| _{[0,T]})\| _{L^2[0,T]}$ and $\| w\| _{C_{\Lambda (T)}}\doteq \| w(\cdot |_{[0,T]})\| _{C[0,T]}$. If $W(x_1,x_2)=w(x_2)$ where $w\in L^2_{\Lambda (T)}({\mathbb R};{\mathbb R})$, then for the function $W$ one can choose a non-degenerate period lattice $\Lambda \subset {\mathbb R}^2$ with basis vectors $E^1=\alpha e_1$ and $E^2=Te_2$ where $\alpha $ is an arbitrary number. Then $W_Y=0$ if $Y_1\neq 0$, and
$$
W_Y=w_{Y_2}\doteq T^{-1}\int _{\, 0}^{\, T}w(t)\, e^{-iY_2t}\, dt
$$
if $Y_1=0$, where $Y_j=(Y,e_j)$, $j=1,2$, and $Y_2=2\pi T^{-1}n$, $n\in {\mathbb Z}$. 
\vskip 0.3cm

{\bf 0.2. Some known results about the spectrum of operator (0.1).} In [5], the Landau Hamiltonian $\widehat H_B+V$ was considered for perodic point potentials $V$. It was proved that there are eigenvalues (of infinite multiplicity) at the Landau levels $\lambda =(2m+1)B$, $m\in {\mathbb Z}_+$, if the period lattice $\Lambda $ is monatomic and $1<\eta \in {\mathbb Q}$ (and the spectrum also has an absolutely continuous component). 

Periodic electric potentials $V\in L^2_{\Lambda }({\mathbb R}^2;{\mathbb R})$ are $-\Delta $-bounded\footnote{$\Delta ={\partial }^2/{\partial }x_1^2+{\partial }^2/{\partial }x_2^2$ is the Laplacian.} with relative bound zero (and therefore bounded with bound zero relative to the operators $\widehat H_B$) [6, Theorem XIII.96]. If $V\in L^2_{\Lambda }({\mathbb R}^2;{\mathbb R})$ and $0<\eta \in {\mathbb Q}$, then the magnetic Floquet--Bloch theory applies 
(when replacing the period lattice $\Lambda $ by a period lattice with basis vectors $QE^1$ and $E^2$, where $Q\in {\mathbb N}$, such that $\eta \in {\mathbb N}$), hence it follows that the spectrum of operator (0.1) has no singular continuous component for all potentials $V\in L^2_{\Lambda }({\mathbb R}^2;{\mathbb R})$ [6,7,8]. In this case, the absence of eigenvalues means the absolute continuity of the spectrum of operator (0.1).

It was shown in [2] that, for any period lattice $\Lambda \subset {\mathbb R}^2$, in the Banach space $(C_{\Lambda }({\mathbb R}^2;{\mathbb R}),\| \cdot \| _{C_{\Lambda }})$, there exits a dense $G_{\delta }$-set\footnote{The complement of the set ${\mathcal O}$ is a set of first Baire category.} ${\mathcal O}$ such that, for any potential $V\in {\mathcal O}$ and for any homogeneous magnetic field with the flux $0<\eta \in {\mathbb Q}$, the spectrum of operator (0.1) is absolutely continuous. A similar result for potentials $V\in L^p_{\Lambda }({\mathbb R}^2;{\mathbb R})$, $p>1$, was obtained in [9], and for potentials from the spaces $L^{\infty }_{\Lambda }({\mathbb R}^2;{\mathbb R})$, $C^n_{\Lambda }({\mathbb R}^2;{\mathbb R})$, $n\in {\mathbb N}$,  and $H^s_{\Lambda }({\mathbb R}^2;{\mathbb R})$, $s>0$, similar results were obtained in [10]. In [10], the absolute continuity of the spectrum of operator (0.1) was also proved for potentials $C^n_{\Lambda }({\mathbb R}^2;{\mathbb R})$, $n\in {\mathbb N}$, if $0<\eta \in {\mathbb Q}$ and
$$
\varlimsup\limits_{2\pi \Lambda ^*\, \ni \, Y\, \to \infty }|Y|^{\, n+1}\biggl( |V_Y|-\sum\limits_{Y^{\prime }\, \in \, 2\pi \Lambda \, \backslash \, \{ Y\} }|V_{Y^{\prime }}|\, \exp \, \biggl( -\frac {|Y^{\prime }-Y|^2}{4B}\biggr) \biggr) =+\infty
$$
(for $n=0$ also see [11]).
\vskip 0.3cm

{\bf Theorem 1} (see [12]). {\it For any nonconstant periodic electric potential $V\in L^2_{\Lambda }({\mathbb R}^2;{\mathbb R})$ and any homogeneous magnetic field with the flux $\eta \in \{ Q^{-1}:Q\in {\mathbb N}\} $, the spectrum of operator $\mathrm (0.1)$ has no eigenvalues outside the set $\{ (2m+1)B+V_0:m\in {\mathbb N}\} $. Moreover, if $V\in L^2_{\Lambda }({\mathbb R}^2;{\mathbb R})\backslash C^{\infty }_{\Lambda }({\mathbb R}^2;{\mathbb R})$, then the spectrum of operator $\mathrm (0.1)$ is absolutely continuous.}
\vskip 0.3cm

{\bf Theorem 2} (see [13]). {\it For any nonconstant trigonometric polynomial $V:{\mathbb R}^2\to {\mathbb R}$ $\mathrm ($with a period lattice $\Lambda $$\mathrm )$ and any homogeneous magnetic field with the flux $\eta \in \{ Q^{-1}:Q\in {\mathbb N}\} $, the spectrum of operator $\mathrm (0.1)$ is absolutely continuous.}
\vskip 0.3cm

If $V(x_1,x_2)=v(x_2)$ where $v\in L^2_{\Lambda (T)}({\mathbb R};{\mathbb R})$, $T>0$, then one can choose the period lattice $\Lambda $ for the potential $V$ with basis vectors $E^1=2\pi T^{-1}B^{-1}e_1$ and $E^2=Te_2$ for which $\eta =1$. Therefore, the following theorem is a direct consequence of Theorem 1 and Theorem 2.
\vskip 0.3cm

{\bf Theorem 3.} {\it Let $V(x_1,x_2)=v(x_2)$, $x\in {\mathbb R}^2$, where $v$ is a nonconstant function from the space $L^2_{\Lambda (T)}({\mathbb R};{\mathbb R})$, $T>0$. Then the spectrum of the operator $\widehat H_B+V$ has no eigenvalues outside the set $\{ (2m+1)B+v_0:m\in {\mathbb N}\} $. Moreover, if $v\in L^2_{\Lambda (T)}({\mathbb R};{\mathbb R})\backslash C^{\infty }_{\Lambda (T)}({\mathbb R};{\mathbb R})$ or the function $v$ is a trigonometric polynomial, then the spectrum of the operator $\widehat H_B+V$ is absolutely continuous.}
\vskip 0.3cm

{\bf 0.3. Definition of functions ${\mathcal W}^{(j)}$ and the main theorem.} Let $C_b({\mathbb R}^2;{\mathbb R})$ be the space of continuous bounded functions $W:{\mathbb R}^2\to {\mathbb R}$, and let $C_b^{\infty }({\mathbb R}^2;{\mathbb R})$ be the space of functions $W\in C^{\infty }({\mathbb R}^2;{\mathbb R}) \cap C_b({\mathbb R}^2;{\mathbb R})$  
such that all partial derivatives ${\partial }^{n_1+n_2}/{\partial }x_1^{n_1}{\partial }x_2^{n_2}$, $n_1,n_2 \in {\mathbb Z}_+$, belong to the space
$C_b({\mathbb R}^2;{\mathbb R})$. By ${\mathfrak L}_m(B)$, $m\in {\mathbb N}$, we denote the set of ordered sets $\{ {\mathcal W}^{(j)}\} $ of functions ${\mathcal W}^{(j)}\in C_b^{\infty }({\mathbb R}^2;{\mathbb R})$, $j=0,\dots , m-1$, provided that the following conditions are fulfilled:

(a)\ ${\mathcal W}^{(j)}(x)<0$ for all $x\in {\mathbb R}^2$,

(a1)\ $1/{\mathcal W}^{(j)}(\cdot )\in C_b({\mathbb R}^2;{\mathbb R})$ (then $1/{\mathcal W}^{(j)}(\cdot )\in C_b^{\infty }({\mathbb R}^2;{\mathbb R})$),

(b)\ for all $j=1,\dots , m-1$
$$
{\mathcal W}^{(j)}={\mathcal W}^{(0)}+2Bj-\Delta \ln \, (|{\mathcal W}^{(0)}|^j\, |{\mathcal W}^{(1)}|^{j-1}\dots |{\mathcal W}^{(j-1)}|),
$$

(c)\ ${\mathcal W}^{(m)}\doteq {\mathcal W}^{(0)}+2Bm-\Delta \ln \, (|{\mathcal W}^{(0)}|^m\, |{\mathcal W}^{(1)}|^{m-1}\dots |{\mathcal W}^{(m-1)}|)\equiv 0$.
\vskip 0.1cm

the following theorem is the mail result of this article.
\vskip 0.3cm

{\bf Theorem 4.} {\it Let $\{ {\mathcal W}^{(j)}\} \in {\mathfrak L}_m(B)$, $m\in {\mathbb N}$. Then the Landau Hamiltonian $\widehat H_B+V$ $\mathrm ($with a homogeneous magnetic field of strength $B>0$$\mathrm )$ has the eigenvalue $\lambda =(2m+1)B$ $\mathrm ($of infinite multiplicity$\mathrm )$ for the electric potential $V={\mathcal W}^{(0)}+2Bm$.} 
\vskip 0.3cm

For non-degenerate lattice $\Lambda \subset {\mathbb R}^2$ let ${\mathfrak L}_m^{\Lambda }(B)$, $m\in {\mathbb N}$, be the set of ordered sets $\{ {\mathcal W}^{(j)}\} \in {\mathfrak L}_m(B)$ such that $\{ {\mathcal W}^{(j)}\} \in C_{\Lambda }^{\infty }({\mathbb R}^2;{\mathbb R})$ for all $j=0, \dots , m-1$.

If $\{ {\mathcal W}^{(j)}\} \in {\mathfrak L}_m^{\Lambda }(B)$, $m\in {\mathbb N}$, it follows from the condition (c) that ${\mathcal W}^{(0)}_0=-2Bm$. Then the condition (b) implies that the functions ${\mathcal W}^{(j)}$, $j=0, \dots , m-1$, have mean values ${\mathcal W}^{(j)}_0=-2B(m-j)$.
\vskip 0.3cm

{\bf Theorem 5.} {\it Let $\Lambda \subset {\mathbb R}^2$ be any non-degenerate lattice and let $\{ {\mathcal W}^{(j)}\} \in {\mathfrak L}_m^{\Lambda }(B)$, $m\in {\mathbb N}$. Then the Landau Hamiltonian $\widehat H_B+V$ $\mathrm ($with a homogeneous magnetic field of strength $B>0$$\mathrm )$ has the eigenvalue $\lambda =(2m+1)B$ $\mathrm ($of infinite multiplicity$\mathrm )$ for the electric potential $V={\mathcal W}^{(0)}+2Bm$ with zero mean value $V_0=0$.} 
\vskip 0.3cm

Theorem 5 is a direct consequence of Theorem 4. A particular case of Theorem 5 was proved in [4] for 
$\{ {\mathcal W}^{(j)}\} \in {\mathfrak L}_1^{\Lambda }(B)$ and $\eta \in {\mathbb Q}$.
\vskip 0.3cm

{\bf 0.4. Existence of functions ${\mathcal W}^{(j)}$.} For functions ${\mathcal W}^{(j)}\in C_b^{\infty }({\mathbb R}^2;{\mathbb R})$, $j=0, \dots , m-1$, conditions (a) and (a1) hold iff
$$
{\mathcal W}^{(j-1)}=-2B(m-j+1)e^{u_j} \eqno(0.2)
$$
where $u_j\in C_b^{\infty }({\mathbb R}^2;{\mathbb R})$, $j=1, \dots , m$. Conditions (b) and (c) are fulfilled iff
$$
-\Delta u_1=2B(e^{u_1}-1) \eqno(0.3)
$$
for $m=1$, and
$$
-\Delta u_1=2Bm(e^{u_1}-1)-2B(m-1)(e^{u_2}-1),
$$ $$
-j\Delta u_1-(j-1)\Delta u_2- \dots -\Delta u_j
=2Bm(e^{u_1}-1)-2B(m-j)(e^{u_{j+1}}-1), \quad j=2, \dots , m-1,
$$ $$
-m\Delta u_1-(m-1)\Delta u_2- \dots -\Delta u_m
=2Bm(e^{u_1}-1) \eqno(0.4)
$$
for $m\geqslant 2$. Furthermore, the inclusions $\{ {\mathcal W}^{(j)}\} \in {\mathfrak L}_m^{\Lambda }(B)$, $m\in {\mathbb N}$, hold iff equality (0.2) and equations (0.3), (0.4) hold for functions $u_j\in C_{\Lambda }^{\infty }({\mathbb R}^2;{\mathbb R})$, $j=1, \dots , m$.

If $m=1$ and $u_1(x_1,x_2)=u(x_2)$, then equation (0.3) becomes an ordinary differential equation
$$
-\frac {d^2u}{dt^2}=2B(e^u-1), \quad t\in {\mathbb R}. \eqno(0.5)
$$
All solutions of equation (0.5) are analytic and periodic.
\vskip 0.3cm

{\bf Lemma 1.} {\it Let $u(\alpha ;\cdot )$, $\alpha >0$, be a solution of equation ${\mathrm (0.5)}$ such that $u(\alpha ;0)=0$, $\frac {du(\alpha ;t)}{dt}\big| _{t=0}=\alpha $, and let $T_{\alpha }>0$ be the minimal period of the function $u(\alpha ;\cdot )$. Then the function $(0,+\infty )\ni \alpha \mapsto T_{\alpha }$ is continuous, $T_{\alpha }\to 2\pi (2B)^{-1/2}$ as $\alpha \to +0$ and $T_{\alpha }\to +\infty $ as $\alpha \to +\infty $.} 
\vskip 0.3cm

The following theorem is a consequence of Theorem 5.
\vskip 0.3cm

{\bf Theorem 6} (see [4]). {\it For any homogeneous magnetic field of a strength $B>0$ and any solution of equation ${\mathrm (0.5)}$, the Landau Hamiltonian $\widehat H_B+V$ has the eigenvalue $\lambda =3B$ 
$\mathrm ($of infinite multiplicity$\mathrm )$ for the electric potential $V=2B(1-e^{u(x_2)})$ 
$\mathrm ($with zero mean value $V_0=0$$\mathrm )$.} 
\vskip 0.3cm

If we subtract the ($j-1$)-th equation of the system (4) from the $j$-th equation, $j=2, \dots , m$, we get the system of equations
$$
-\Delta u_1=2Bm(e^{u_1}-1)-2B(m-1)(e^{u_2}-1),
$$ $$
-\Delta u_1-\Delta u_2- \dots -\Delta u_j
=2B(m-j+1)(e^{u_j}-1)-2B(m-j)(e^{u_{j+1}}-1), \quad j=2, \dots , m-1,
$$ $$
-\Delta u_1-\Delta u_2- \dots -\Delta u_m
=2B(e^{u_m}-1). \eqno(0.6)
$$
If we do the same operations for the system (0.6), we get the system of equations
$$
-\Delta u_j=B\sum\limits_{\mu \, =\, 1}^mC^{(m)}_{j\mu }(e^{u_{\mu }}-1), \quad j=1, \dots , m,
\eqno(0.7)
$$
which is equivalent to the equation (0.3) for $m=1$ and is equivalent to the system of equations (0.4) for $m\geqslant 2$, where $C^{(m)}_{j\mu }$ are elements of a ($m\times m$)-matrix $\widehat C^{(m)}$, $m\in {\mathbb N}$; $C^{(1)}_{11}=2$ if $m=1$, and for $m\geqslant 2$ the following elements of the matrix $\widehat C^{(m)}$ are chosen: $C^{(m)}_{11}=2m$, $C^{(m)}_{jj}=4(m-j+1)$ if $j=2, \dots , m$, $C^{(m)}_{j,j-1}=-2(m-j+2)$ if $j=2, \dots , m$, $C^{(m)}_{j,j+1}=-2(m-j)$ if $j=1, \dots , m-1$, and $C^{(m)}_{j\mu }=0$ otherwise.
\vskip 0.3cm

{\bf Lemma 2.} {\it All eigenvalues ${\lambda }^{(m)}_j$, $j=1, \dots , m$, of the matrix $\widehat C^{(m)}$ are real and $0<{\lambda }^{(m)}_1< \dots <{\lambda }^{(m)}_m$. Moreover, for any eigenvector\footnote{The sign ``T'' represents the transposition of a row vector.} $a=(a_1, \dots , a_m)^T$, the inequality $a_1\neq 0$ holds.} 
\vskip 0.3cm

Let $B_0>0$. For a given $m\in {\mathbb N}$, let
denote by ${\lambda }^{(m)}$ any eigenvalue of the matrix $\widehat C^{(m)}$ such that the following condition is fulfilled:
$$
n^2{\lambda }^{(m)}\neq {\lambda }^{(m)}_j\ \ for\ all\ \ n\in {\mathbb N}\backslash \{ 1\} \ \ and\ all\ \ {\lambda }^{(m)}_j>{\lambda }^{(m)}. \eqno(0.8)
$$
Let $\omega \doteq \sqrt {B_0{\lambda }^{(m)}}$. By $a=(a_1, \dots , a_m)^T\in {\mathbb R}^m$ we denote the eigenvector of the matrix $\widehat C^{(m)}$ with the eigenvalue ${\lambda }^{(m)}$, $\| a\| _{{\mathbb R}^m}=\bigl( \, \sum a_j^2\, \bigr) ^{1/2}=1$ and $a_1>0$. Let $T=2\pi {\omega }^{-1}$, $\Lambda (T)=T{\mathbb Z}$, and ${\Lambda }^*=T^{-1}{\mathbb Z}=(2\pi )^{-1}{\omega }{\mathbb Z}$.

Let $L^{2,\, even}_{\Lambda (T)}({\mathbb R};{\mathcal B})$, $C^{\, even}_{\Lambda (T)}({\mathbb R};{\mathcal B})$, and $C^{\infty ,\, even}_{\Lambda (T)}({\mathbb R};{\mathcal B})$, where ${\mathcal B}={\mathbb R}$ or ${\mathcal B}={\mathbb C}$, be the subspaces of even T-periodic functions from the spaces $L^2_{\Lambda (T)}({\mathbb R};{\mathcal B})$,  $C_{\Lambda (T)}({\mathbb R};{\mathcal B})$, and $C^{\infty }_{\Lambda (T)}({\mathbb R};{\mathcal B})$, respectively.

If $u_j(x_1,x_2)=v_j(x_2)$ where $v_j(\cdot )\in C^{\infty }_{\Lambda (T)}({\mathbb R};{\mathbb R})$, then the functions $u_j$ satisfy system (0.7) iff the functions $v_j$ satisfy the system of ordinary differential equations
$$
-\, \frac {d^2v_j}{dt^2}= B\, \sum\limits_{\mu \, =\, 1}^mC^{(m)}_{j\mu }\bigl( e^{v_{\mu }}-1\bigr) ,\quad j=1, \dots , m. \eqno(0.9)
$$

For $B=B_0$, the T-periodic functions $\widetilde v_j=2\varepsilon a_j \cos \, {\omega }t$, $j=1, \dots , m$, $\varepsilon \in {\mathbb R}$, are solutions of the linearized system (for system (0.9))
$$
-\, \frac {d^2\widetilde v_j}{dt^2}= B_0\, \sum\limits_{\mu \, =\, 1}^mC^{(m)}_{j\mu }\,\widetilde v_{\mu } ,\quad j=1, \dots , m.
$$
In the following theorem the functions $\widetilde v_j$ are the leader terms in expansion in degrees of a small parameter $\varepsilon \in {\mathbb R}$ of solutions $v_j$ of nonlinear system (0.9). The strength of a homogeneous magnetic field $B>0$ also depends on the parameter $\varepsilon $.
\vskip 0.3cm

{\bf Theorem 7.} {\it Let $B_0>0$ and let $m\in {\mathbb N}$. Suppose that the condition $\mathrm  (0.8)$ for the eigenvalue ${\lambda }^{(m)}$ of the matrix $\widehat C^{(m)}$ is fulfilled. Then there exist a   
number $r>0$ and analytic functions\footnote{the number $r>0$ and the function ${\tau }(\cdot )$ do not depend on $B_0$.} $(-r,r)\ni \varepsilon \mapsto {\tau }(\varepsilon )\in {\mathbb R}$ and $(-r,r)\ni \varepsilon \mapsto {\mathfrak w}_j(\varepsilon ;\cdot )\in (C_{\Lambda (T)}({\mathbb R};{\mathbb R}), \| \cdot \| _{C_{\Lambda (T)}})$, $j=1, \dots , m$, such that, for all $\varepsilon \in (-r,r)$ the functions ${\mathfrak w}_j(\varepsilon ;\cdot )$ belong to the subspace $C^{\infty ,\, even}_{\Lambda (T)}({\mathbb R};{\mathbb R})$, $1+{\varepsilon }^2{\tau }(\varepsilon )>0$, and the functions
$$
{\mathbb R}\ni t\mapsto v_j(t)=v_j(\varepsilon ;t)\doteq 2\varepsilon a_j \cos \, {\omega }t+{\varepsilon }^2{\mathfrak w}_j(\varepsilon ;t),\quad j=1,\dots , m, \eqno(0.10)
$$
are solutions of the system
$$
-\, \frac {d^2v_j}{dt^2}= B_0(1+{\varepsilon }^2{\tau }(\varepsilon ))\, \sum\limits_{\mu \, =\, 1}^mC^{(m)}_{j\mu }\bigl( e^{v_{\mu }}-1\bigr) ,\quad j=1, \dots , m. \eqno(0.11)
$$}
\vskip 0.3cm

The following theorem is a consequence of Theorem 5, Theorem 7, and equality (0.2).
\vskip 0.3cm

{\bf Theorem 8.} {\it Let $B_0>0$ and let $m\in {\mathbb N}$. Then $\mathrm ($assuming the conditions of Theorem 7$\mathrm )$ for any $\varepsilon \in (-r,r)$ and for a homogeneous magnetic field with the strength $B_0(1+{\varepsilon }^2{\tau }(\varepsilon ))>0$, the Landau Hamiltonian $\widehat H_B+V$ has the eigenvalue $\lambda =(2m+1)B$ $\mathrm ($of infinite multiplicity$\mathrm )$ for the electric potential ${\mathbb R}^2\ni x\mapsto V(x)=2Bm(1-e^{u_1(\varepsilon ;x_2)})$ which is $T$-periodic in $x_2$ and has zero mean value $V_0=0$.}
\vskip 0.3cm

In Section 1 we prove Theorem 4. Some properties of solutions of equation (5) are considered in Section 2 where Lemma 1 is proved. Statements about eigenvalues and eigenvectors of the matrices $\widehat C^{(m)}$, $m\in {\mathbb N}$, are collected in Section 3. In Section 4 we prove Theorem 7.

\section{Some auxiliary results and the proof of Theorem 4}

Let $\| \cdot \| _{L^2}$ and $(\cdot ,\cdot )_{L^2}$ be the norm and the inner product in the Hilbert space $L^2({\mathbb R}^2;{\mathbb C})$. The inner product $(\cdot ,\cdot )_{L^2}$ is supposed to be linear in the second argument. Define the operators
$$
\widehat Z_{\mp }=-i\, \frac {\partial }{\partial x_1}\pm \biggl( -i\, \frac {\partial }{\partial x_2}-Bx_1\biggr) 
$$
acting on $L^2({\mathbb R}^2;{\mathbb C})$ with domains $D(\widehat Z_{\mp })=\bigl\{ \Phi \in L^2({\mathbb R}^2;{\mathbb C})\cap H^1_{\mathrm {loc}}({\mathbb R}^2;{\mathbb C}): \widehat Z_{\mp }\Phi \in L^2({\mathbb R}^2;{\mathbb C})\bigr\} $.
We have
$$
\widehat H_B=\widehat Z_+\widehat Z_-+B=\widehat Z_-\widehat Z_+-B.
$$

Let ${\mathcal H}^{(m)}_B$, $m\in {\mathbb Z}_+$, be subspaces of eigenfunctions of the operator $\widehat H_B$ with eigenvalues $\lambda =(2m+1)B$. The linear subspaces ${\mathcal H}^{(m)}_B\cap {\mathcal S}({\mathbb R}^2;{\mathbb C})$ are dense in ${\mathcal H}^{(m)}_B$ (where ${\mathcal S}({\mathbb R}^2;{\mathbb C})$ is the Schwartz space) and ${\mathcal H}^{(m)}_B\subset C^{\infty }({\mathbb R}^2;{\mathbb C})$, $m\in {\mathbb Z}_+$. By $\widehat P^{(m)}$ we denote the orthogonal projection in $L^2({\mathbb R}^2;{\mathbb C})$ onto the subspace ${\mathcal H}^{(m)}_B$.

If $\Phi \in {\mathcal H}^{(m)}_B$, $m\in {\mathbb Z}_+$, then $\widehat Z_+\Phi \in {\mathcal H}^{(m+1)}_B$ and $\| \widehat Z_+\Phi \| _{L^2}=\sqrt {2B(m+1)}\, \| \Phi \| _{L^2}$. If $\Phi \in {\mathcal H}^{(m)}_B$, $m\in {\mathbb N}$, then $\widehat Z_-\Phi \in {\mathcal H}^{(m-1)}_B$ and $\| \widehat Z_-\Phi \| _{L^2}=\sqrt {2Bm}\, \| \Phi \| _{L^2}$. For all $\Phi \in {\mathcal H}^{(0)}_B$, the equality $\widehat Z_-\Phi =0$ holds.

Let denote 
$$
{\mathcal H}^{\infty }_B\doteq \biggl\{ \Phi \in 
L^2({\mathbb R}^2;{\mathbb C}): \sum\limits_{m\, \in \, {\mathbb Z}_+}m^p\, \| \widehat P^{(m)}
\Phi \| ^2_{L^2}<+\infty \ \ for\ all\ \ p>0\biggr\} ;
$$
${\mathcal S}({\mathbb R}^2;{\mathbb C})\subset 
{\mathcal H}^{\infty }_B\subset C^{\infty }({\mathbb R}^2;{\mathbb C})$. For any function $\Phi \in {\mathcal H}^{\infty }_B$ there exists a sequence of functions ${\Phi }_j\in {\mathcal S}({\mathbb R}^2;{\mathbb C})$, $j\in {\mathbb N}$, such that $\| \widehat Z_+^n(\Phi -\Phi _j\| _{L^2}\to 0$ as $j\to +\infty $ for all $n\in {\mathbb Z}_+$. The last assertion and the closedness of the operators $\widehat Z_{\mp }$ imply that the equalities
$$
\widehat Z_{\mp }{\mathcal W}\Phi = -i\, \biggl( \frac {\partial {\mathcal W}}{\partial x_1}\pm i\, \frac {\partial {\mathcal W}}{\partial x_2}\biggr) +{\mathcal W}\widehat Z_{\mp }\Phi \eqno(1.1)
$$
hold for all functions $\Phi \in {\mathcal H}^{\infty }_B$ and ${\mathcal W}\in C^{\infty }_b({\mathbb R}^2;{\mathbb C})$.

Right inverses of the operator $\widehat Z_-$ are bounded linear operators but are not uniquely defined. Let $\widehat Z_-^{-1}$ be the operator such that
$$
\widehat Z_-^{-1}\Phi =\sum\limits_{m\, \in \, {\mathbb Z}_+}\bigl( 2B(m+1)\bigr) ^{-1}\widehat Z_+\widehat P^{(m)}\Phi ,\quad \Phi \in L^2({\mathbb R}^2;{\mathbb C}).
$$
If $\Phi \in D(\widehat Z_-)$, then
$$
\widehat Z_-^{-1}\widehat Z_-\Phi =\Phi -\widehat P^{(0)}\Phi . \eqno(1.2)
$$

For any function $\Phi \in {\mathcal H}^{\infty }_B$, we have $\widehat Z_{\mp }\Phi \in {\mathcal H}^{\infty }_B$ and $\widehat Z_-^{-1}\Phi \in {\mathcal H}^{\infty }_B$.
\vskip 0.3cm

{\bf Lemma 3.} {\it If $\Phi \in {\mathcal H}^{\infty }_B$ and ${\mathcal W}\in C^{\infty }_b({\mathbb R}^2;{\mathbb C})$, then ${\mathcal W}\Phi \in {\mathcal H}^{\infty }_B$.}
\vskip 0.3cm

{\it Proof}. Let $\Phi ^{\prime }\in {\mathcal H}^{(m)}_B$, $m\in {\mathbb Z}_+$, and $q\in {\mathbb N}$. Because
$$
(\Phi ^{\prime },{\mathcal W}\Phi )_{L^2}=\bigl( (2B)^q(m+1)\dots (m+q)\bigr) ^{-1}(\Phi 
^{\prime },\widehat Z_-^q\widehat Z_+^q{\mathcal W}\Phi )_{L^2},
$$
it follows from (1.1) that
$$
\| \widehat P^{(m)}{\mathcal W}\Phi \| _{L^2}\leqslant C\bigl( (2B)^q(m+1)\dots (m+q)\bigr) ^{-1}
$$
where $C=C(B,q;{\mathcal W},\Phi )>0$. Hence, ${\mathcal W}\Phi \in {\mathcal H}^{\infty }_B$.
\hfill $\square $
\vskip 0.3cm

Let $\{ {\mathcal W}^{(j)}\} \in {\mathfrak L}_m(B)$, $m\in {\mathbb N}$, and let
$$
\widehat {\mathcal B}^{(0)}=\widehat Z_-^{-1}(\widehat Z_+\widehat Z_-+{\mathcal W}^{(0)}). \eqno(1.3)
$$
Sequentially for $j=1, \dots , m$ we define the operators
$$
\widehat {\mathcal B}^{(j)}\doteq ({\mathcal W}^{(j-1)})^{-1}\widehat Z_-\widehat {\mathcal B}^{(j-1)}\widehat Z_-^{-1}{\mathcal W}^{(j-1)}. \eqno(1.4)
$$
All these operators can be considered as operators acting on ${\mathcal H}^{\infty }_B$.
\vskip 0.3cm

{\bf Lemma 4.} {\it For all $j=1, \dots , m$,
$$
\widehat {\mathcal B}^{(j)}=\widehat Z_++\widehat Z_-^{-1}{\mathcal W}^{(j-1)}-i\, \biggl( \biggl( \frac {\partial }{\partial x_1}- i\, \frac {\partial }{\partial x_2}\biggr) \ln \, |{\mathcal W}^{(0)}\dots {\mathcal W}^{(j-1)}|\biggr) , \eqno(1.5)
$$
and for all $j=0,1, \dots , m$ and $\Psi \in {\mathcal H}_B^{(0)}$,
$$
\widehat Z_-\widehat {\mathcal B}^{(j)}\Psi={\mathcal W}^{(j)}\Psi . \eqno (1.6)
$$}
\vskip 0.3cm

{\it Proof}. We have
$$
\widehat {\mathcal B}^{(1)}\doteq ({\mathcal W}^{(0)})^{-1}\widehat Z_-\widehat {\mathcal B}^{(0)}\widehat Z_-^{-1}{\mathcal W}^{(0)}=({\mathcal W}^{(0)})^{-1}(\widehat Z_++{\mathcal W}^{(0)}\widehat Z_-^{-1}){\mathcal W}^{(0)}
$$ $$
=\, \widehat Z_++\widehat Z_-^{-1}{\mathcal W}^{(0)}-i\, \biggl( \biggl( \frac {\partial }{\partial x_1}- i\, \frac {\partial }{\partial x_2}\biggr) \ln \, |{\mathcal W}^{(0)}|\biggr) .
$$
Now, assume that (1.5) holds for some $j\in {\mathbb N}$, $j<m$. Then
$$
\widehat {\mathcal B}^{(j+1)}= ({\mathcal W}^{(j)})^{-1}\widehat Z_-\widehat {\mathcal B}^{(j)}\widehat Z_-^{-1}{\mathcal W}^{(j)}
$$ $$
=\, ({\mathcal W}^{(j)})^{-1}(\widehat Z_++(2B+
{\mathcal W}^{(j-1)}-\Delta \ln \, |{\mathcal W}^{(0)}\dots {\mathcal W}^{(j-1)}|)\widehat Z_-^{-1}){\mathcal W}^{(j)}
$$ $$
-i\, \bigg( \biggl( \frac {\partial }{\partial x_1}-i\, \frac {\partial }{\partial x_2}\biggr) \ln \, |{\mathcal W}^{(0)}\dots {\mathcal W}^{(j-1)}|\biggr)
$$ $$
=\, \widehat Z_+-i\, \bigg( \biggl( \frac {\partial }{\partial x_1}-i\, \frac {\partial }{\partial x_2}\biggr) \ln \, |{\mathcal W}^{(j)}|\biggr) +\widehat Z_-^{-1}{\mathcal W}^{(j)}
$$ $$
-i\, \bigg( \biggl( \frac {\partial }{\partial x_1}-i\, \frac {\partial }{\partial x_2}\biggr) \ln \, |{\mathcal W}^{(0)}\dots {\mathcal W}^{(j-1)}|\biggr)
$$ $$
=\, \widehat Z_++\widehat Z_-^{-1}{\mathcal W}^{(j)}-i\, \biggl( \biggl( \frac {\partial }{\partial x_1}- i\, \frac {\partial }{\partial x_2}\biggr) \ln \, |{\mathcal W}^{(0)}\dots {\mathcal W}^{(j-1)}{\mathcal W}^{(j)}|\biggr) .
$$
We have shown that (1.5) is fulfilled for $j+1$, therefore (1.5) follows by induction for all $j=1, \dots , m$. By (1.3), the equality
$$
\widehat Z_-\widehat {\mathcal B}^{(0)}\Psi ={\mathcal W}^{(0)}\Psi
$$
holds for all $\Psi \in {\mathcal H}^{(0)}_B$. Hence, using (1.5), for all $j=1, \dots , m$ and $\Psi \in {\mathcal H}^{(0)}_B$, we obtain
$$
\widehat Z_-\widehat {\mathcal B}^{(j)}\Psi =
(2B+{\mathcal W}^{(j-1)}-\Delta \ln \, |{\mathcal W}^{(0)}\dots {\mathcal W}^{(j-1)}|)\Psi = {\mathcal W}^{(j)}\Psi .
$$
Lemma 4 is proved.
\vskip 0.3cm

THE PROOF OF THEOREM 4. It follows from (1.6) that, for all $j=1, \dots , m$ and for any function $\Psi \in {\mathcal H}^{(0)}_B$, there is a function $\Psi _{j,1} \in {\mathcal H}^{(0)}_B$ such that
$$
\widehat {\mathcal B}^{(j)}\Psi =\widehat Z_-^{-1}{\mathcal W}^{(j)}\Psi +\Psi _{j,1} . \eqno(1.7)
$$
For $j=1, \dots , m$, under sequential action of the operators $\widehat Z_-^{-1}{\mathcal W}^{(j-1)}, \dots , \widehat Z_-^{-1}{\mathcal W}^{(0)}$ on left-hand side and right-hand side of equality (1.7), using (1.2), (1.3), and (1.4), we get the equations
$$
\widehat {\mathcal B}^{(0)}\widehat Z_-^{-1}{\mathcal W}^{(0)}\dots \widehat Z_-^{-1}{\mathcal W}^{(j-1)}\Psi
$$ $$
=\, \widehat Z_-^{-1}{\mathcal W}^{(0)}\dots \widehat Z_-^{-1}{\mathcal W}^{(j-1)}\widehat Z_-^{-1}{\mathcal W}^{(j)}\Psi +\sum\limits_{s\, =\, 1}^j\widehat Z_-^{-1}{\mathcal W}^{(0)}\dots \widehat Z_-^{-1}{\mathcal W}^{(j-s)}\Psi _{j,s}
\eqno(1.8)
$$
where $\Psi _{j,s}\in {\mathcal H}^{(0)}_B$, $s=1, \dots , j$. Since ${\mathcal W}^{(m)}\equiv 0$, for all $\Psi \in {\mathcal H}^{(0)}_B$, we have
$$
\widehat {\mathcal B}^{(0)}\widehat Z_-^{-1}{\mathcal W}^{(0)}\dots \widehat Z_-^{-1}{\mathcal W}^{(m-1)}\Psi =\sum\limits_{s\, =\, 1}^m\widehat Z_-^{-1}{\mathcal W}^{(0)}\dots \widehat Z_-^{-1}{\mathcal W}^{(m-s)}\Psi _{m,s}. 
\eqno(1.9)
$$
From (1.8), (1.9) and from the equality $\widehat {\mathcal B}^{(0)}\Psi =\widehat Z_-^{-1}{\mathcal W}^{(0)}\Psi $, it follows that for any function $\Psi \in {\mathcal H}^{(0)}_B$ we can sequentially choose functions $\Psi _0, \dots , \Psi _{m-1}\in {\mathcal H}^{(0)}_B$ (which are uniquely determined) such that, for the function
$$
\Phi (\Psi )\doteq \widehat Z_-^{-1}{\mathcal W}^{(0)}\dots \widehat Z_-^{-1}{\mathcal W}^{(m-1)}\Psi +\sum\limits_{s\, =\, 0}^{m-2}\widehat Z_-^{-1}{\mathcal W}^{(0)}\dots \widehat Z_-^{-1}{\mathcal W}^{(m-2-s)}\Psi _s+\Psi _{m-1}\in 
{\mathcal H}^{(0)}_B , \eqno(1.10)
$$
the equality $\widehat {\mathcal B}^{(0)}(\Phi (\Psi ))\equiv 0$ holds (in (1.10) the sum is absent if $m=1$). Using the equality $({\mathcal W}^{(m-1)})^{-1}\widehat Z_-\dots ({\mathcal W}^{(0)})^{-1}\widehat Z_-\Phi (\Psi )=\Psi $, we obtain that $\Phi (\Psi )\not\equiv 0$ if $\Psi \not\equiv 0$. Moreover, for any finite set of linearly independent functions $\Psi \in {\mathcal H}^{(0)}_B$, the functions $\Phi (\Psi )$ also are linearly independent. By (1.3),
$$
(\widehat Z_+\widehat Z_-+{\mathcal W}^{(0)})\, \Phi (\Psi )\equiv 0
$$
for all non-zero functions $\Psi \in {\mathcal H}^{(0)}_B$. This concludes the proof of Theorem 4.

\section{The proof of Lemma 1}

Let $u_+(\alpha )$ and $u_-(\alpha )$ denote the maximum value and the minimum value of the solution $u(\alpha ;\cdot )$, $\alpha >0$, of equation (0.5). Then
$$
e^{u_+(\alpha )}-1-u_+(\alpha )=e^{u_-(\alpha )}-1-u_-(\alpha )=(4B)^{-1}{\alpha }^2
$$
and
$$
\biggl( \frac {du(\alpha ;t)}{dt}\biggr) ^2={\alpha }^2-4B (e^{u(\alpha ;t)}-1-u(\alpha ;t)).
$$
Thus, for the (minimal) periods $T_{\alpha }$ of oscillations of the solutions $u(\alpha ;\cdot )$, we obtain
$$
T_{\alpha }=2\int_{u_-(\alpha )}^{\, u_+(\alpha )}\frac {du}{\sqrt {{\alpha }^2-4B(e^u-1-u)}}=
\alpha B^{-1}\int_{-1}^{\, 1}\frac {\xi }{e^{u_{\xi }}-1}\, \frac {d\xi }
{\sqrt {1-{\xi }^2}} \eqno(2.1)
$$
where $[-1,1]\ni \xi \mapsto u_{\xi }\in {\mathbb R}$ is a monotonically increasing function such that $e^{u_{\xi }}-1-u_{\xi }=(4B)^{-1}{\alpha }^2{\xi }^2$; $u_{-1}=-(4B)^{-1}{\alpha }^2$, $u_0=0$, $u_1=(4B)^{-1}{\alpha }^2$. Equality (2.1) implies that the function $(0,+\infty )\ni \alpha \mapsto T_{\alpha }$ is continuous. If $\alpha \to +0$, then $\xi (e^{u_{\xi }}-1)^{-1}\to 2\sqrt B{\alpha }^{-1}$ uniformly in $\xi \in [-1,1]$.
Therefore\footnote{We have $T_{\alpha }=2\pi (2B)^{-1/2}\bigl( 1+\frac 1{48}\, {\alpha }^2B^{-1}+O({\alpha }^4B^{-2})\bigr) $ as $\alpha \to +0$.}, $T_{\alpha }\to 2\pi (2B)^{-1/2}$ as 
$\alpha \to +0$. On the other hand, $-u_{\xi }> 
(4B)^{-1}{\alpha }^2{\xi }^2$ with $\xi \in [-1,0)$. Therefore, $\xi (e^{u_{\xi }}-1)^{-1}\to -\xi $ as $\alpha \to +\infty $ uniformly in $\xi \in [-1,-1/2]$, and hence\footnote{Moreover, $T_{\alpha }=(1+o(1))\alpha B^{-1}$ as $\alpha \to +\infty $.} $T_{\alpha }\to +\infty $ as $\alpha \to +\infty $.  

\section{Eigenvalues of matrices $\widehat C^{(m)}$, $m\in {\mathbb N}$}

{\bf 3.1. The proof of Lemma 2.} We have $\widehat C^{(m)}=\widehat {\mathcal E}^{(m)}\widehat {\mathcal D}^{(m)}$ where $\widehat {\mathcal E}^{(m)}$ is a symmetric ($m\times m$)-matrix with non-zero elements ${\mathcal E}^{(m)}_{11}=1$ and (for $m\geqslant 2$) ${\mathcal E}^{(m)}_{jj}=2$ if $j=2, \dots , m$, and ${\mathcal E}^{(m)}_{j,\, j+1}={\mathcal E}^{(m)}_{j+1,\, j}=-1$ if $j=1, \dots , m-1$. The 
($m\times m$)-matrix $\widehat {\mathcal D}^{(m)}$ is dioganal with non-zero elements ${\mathcal D}^{(m)}_{jj}=2(m-j+1)$, $j=1, \dots , m$.
\vskip 0.3cm

{\bf Lemma 5.} {\it The matrix $\widehat {\mathcal E}^{(m)}$ is positive definite.}
\vskip 0.3cm

{\it Proof}. Let $D_N$, $N\in {\mathbb N}$, be the determinant of the symmetric ($N\times N$)-matrix ${\widehat F}^{(N)}$ with non-zero elements $F^{(N)}_{jj}=2$, $j=1, \dots , N$, and (for $N\geqslant 2$) $F^{(N)}_{j,\, j+1}=F^{(N)}_{j+1,\, j}=-1$, $j=1, \dots , N-1$. Since $D_1=2$, $D_2=3$, and $D_{N+2}=2D_{N+1}-D_N$ if $N\geqslant 1$, it follows that $D_N=N+1$ for all $N\in {\mathbb N}$. On the other hand, $\det \, \widehat {\mathcal E}^{(m)}=1$ if $m=1,2$, and $\det \, \widehat {\mathcal E}^{(m)}=D_{m-1}-D_{m-2}$ if $m\geqslant 3$. Hence $\det \, \widehat {\mathcal E}^{(m)}=1$ for all $m\in {\mathbb N}$. Thus, all minors $\det \, ({\mathcal E}_{jk})_{j,\, k=n, \dots , N}$, $n\in \{ 1, \dots , N\} $, of the matrix $\widehat {\mathcal E}^{(m)}$ are positive and, therefore, the matrix $\widehat {\mathcal E}^{(m)}$ is positive definite. \hfill $\square $
\vskip 0.3cm

Let $(\cdot ,\cdot )_{{\mathbb C}^m}$ and $\| \cdot \| _{{\mathbb C}^m}$ be the inner product and the norm in ${\mathbb C}^m$; $(z^{(1)},z^{(2)})_{{\mathbb C}^m}=\sum\limits_{j\, =\, 1}^m\overline z^{(1)}_jz^{(2)}_j$, $z^{(\mu )}=(z^{(\mu )}_1, \dots , z^{(\mu )}_m)^T$, $z^{(\mu )}_j\in {\mathbb C}$, $j=1, \dots , m$, $\mu =1,2$ (for real vectors we replace ${\mathbb C}^m$ with ${\mathbb R}^m$
in notation); $\widehat I_m$ is the identity ($m\times m$)-matrix.
\vskip 0.3cm

{\bf Lemma 6.} {\it All eigenvalues of the matrix $\widehat {\mathcal E}^{(m)}$ are real and positive.}
\vskip 0.3cm

{\it Proof}. Let $a$ be an eigenvector of the matrix $\widehat {\mathcal E}^{(m)}$ with an eigenvalue $\lambda \in {\mathbb C}$. Then it follows from Lemma 5 that
$$
0<(\widehat {\mathcal D}^{(m)}a,\widehat {\mathcal E}^{(m)}\widehat {\mathcal D}^{(m)}a)_{{\mathbb C}^m}=(\widehat {\mathcal D}^{(m)}a,\widehat C^{(m)}a)_{{\mathbb C}^m}=
\lambda \, (\widehat {\mathcal D}^{(m)}a,a)_{{\mathbb C}^m} .
$$
On the other hand, $(\widehat {\mathcal D}^{(m)}a,a)_{{\mathbb C}^m}>0$. Hence $\lambda >0$. \hfill $\square $
\vskip 0.3cm 

Since, for any $\lambda \in {\mathbb C}$, the first $m-1$ columns of the matrix $\widehat C^{(m)}-\lambda \widehat I_m$ are linearly independent, the inequality ${\mathrm {Rank}}\, (\widehat C^{(m)}-\lambda \widehat I_m)\geqslant m-1$ holds and, therefore, the following lemma is fulfilled.
\vskip 0.3cm

{\bf Lemma 7.} {\it Geometric multiplicity of eigenvalues of the matrix $\widehat C^{(m)}$ equals} 1.
\vskip 0.3cm

{\bf Lemma 8.} {\it Algebraic multiplicity of eigenvalues of the matrix $\widehat C^{(m)}$ equals} 1.
\vskip 0.3cm

{\it Proof}. Lemma 7 implies that it suffices to show that there are no associated vectors for eigenvectors of the matrix $\widehat C^{(m)}$. We argue by contradiction; suppose that a vector $a^{\prime }$ is an associated vector for an eigenvector $a$ with an eigenvalue $\lambda >0$. Then $\widehat C^{(m)}a^{\prime }=\lambda a^{\prime }+a$ and
$$
0=(\widehat {\mathcal D}^{(m)}a^{\prime },(\widehat C^{(m)}-\lambda \widehat I_m)a)_{{\mathbb C}^m}=(\widehat {\mathcal D}^{(m)}a^{\prime },(\widehat {\mathcal E}^{(m)}
-\lambda (\widehat {\mathcal D}^{(m)})^{-1})\widehat {\mathcal D}^{(m)}a)_{{\mathbb C}^m}
$$ $$
=\, ((\widehat {\mathcal E}^{(m)}
-\lambda (\widehat {\mathcal D}^{(m)})^{-1})\widehat {\mathcal D}^{(m)}a^{\prime },\widehat {\mathcal D}^{(m)}a)_{{\mathbb C}^m}
=\, ((\widehat C^{(m)}-\lambda \widehat I_m)a^{\prime },\widehat {\mathcal D}^{(m)}a)_{{\mathbb C}^m}=(a,\widehat {\mathcal D}^{(m)}a)_{{\mathbb C}^m}>0.
$$
We have arrived at a contradiction; this proves Lemma 8.
\vskip 0.3cm

{\bf Lemma 9.} {\it If $a=(a_1, \dots , a_m)^T$ is an eigenvector of the matrix $\widehat C^{(m)}$, then $a_1\neq 0$.}
\vskip 0.3cm

{\it Proof}. It may be assumed that $m\geqslant 2$. Denote by $a$ an eigenvector of the matrix $\widehat C^{(m)}$ with an eigenvalue $\lambda $. Let $\widehat {\mathcal C}$ be the matrix which is obtained from the matrix $\widehat C^{(m)}-\lambda \widehat I_m$ by deletion of the first column and the last row. Suppose that $a_1=0$, then, for the vector $\widetilde a=(a_2, \dots , a_m)^T$, the vector $\widehat {\mathcal C}\,\widetilde a$ is zero. On the other hand,    
${\mathrm {Rank}}\, \widehat {\mathcal C}=m-1$, hence $a_2=\dots =a_m=0$. Contradiction; Lemma 9 is proved.
\vskip 0.3cm

Now, Lemma 2 follows from Lemma 6, Lemma 8, and Lemma 9.
\vskip 0.3cm

{\bf 3.2. Lemma 10 and its proof.} In what follows, eigenvectors $a=(a_1 , \dots , a_m)^T$ 
of the matrix $\widehat C^{(m)}$ will be chosen from the space ${\mathbb R}^m$. Let $B_0>0$ and let $\omega = \sqrt {B_0\lambda ^{(m)}}$ where $\lambda ^{(m)}\in \{ \lambda ^{(m)}_1, \dots , \lambda ^{(m)}_m\} $. Denote by $a=(a_1 , \dots , a_m)^T\in {\mathbb R}^m$ the eigenvector of the matrix $\widehat C^{(m)}$ with the eigevalue $\lambda ^{(m)}$ such that $\| a\| _{{\mathbb R}^m}=1$ and $a_1>0$. Lemma 8 implies that
$$
\bigl\{ (\omega ^2\widehat I_m-B_0\widehat C^{(m)})b: b\in {\mathbb C}^m \ \, and \ \, (b,a)_{{\mathbb C}^m}=0\bigr\} 
$$
is an ($m-1$)-dimensional linear subspace of ${\mathbb C}^m$; this subspace coincides with  a linear span of the eigenvectors of the matrix $\widehat C^{(m)}$ with eigenvalues $\lambda ^{(m)}_j\neq \lambda ^{(m)}$. Since the eigenvector $a$ does not belong to this subspace, the following lemma is valid.
\vskip 0.3cm

{\bf Lemma 10.} {\it For any vector $f=(f_1, \dots , f_m)^T\in {\mathbb C}^m$, there exist a unique number $\tau \in {\mathbb C}$ and a unique vector $b=(b_1, \dots , b_m)^T\in {\mathbb C}^m$ such that $(b,a)_{{\mathbb C}^m}=0$ and
$$
\bigl( \omega ^2\widehat I_m-B_0\widehat C^{(m)}\bigr) b-\tau {\omega }^2a\, =\, B_0f.
$$
Therefore, for some numbers\footnote{${\tau }_{\mu }=-\frac {a_{\mu }}2\, \bigl( (m-\mu +1)\lambda ^{(m)}(\widehat {\mathcal D}^{(m)}a,a)_{{\mathbb R}^m}\bigr) ^{-1}$, $\mu =1, \dots , m$.} ${\tau }_{\mu }\in {\mathbb R}$ and $R_{\nu \mu }\in {\mathbb R}$, $\nu , \mu =1, \dots , m$, which depend on the choice of the eigenvalue $\lambda ^{(m)}$, we have
$$
\tau =\sum\limits_{\mu }\tau _{\mu }f_{\mu },\quad 
b_{\nu } =\sum\limits_{\mu }R_{\nu \mu }f_{\mu },
\quad \nu =1, \dots , m,
$$
and hence
$$
|\tau |+\sum\limits_{\nu }|b_{\nu }|\, \leqslant \, C\, \sum\limits_{\nu }|f_{\nu }|
$$
where $C=C(m)>0$.}

\section{The proof of Theorem 7}

Let $T=2\pi {\omega }^{-1}$, let ${\mathbb Y}=\{ \omega ,-\omega \} $, and let ${\mathbb Y}^{\prime }\doteq \omega {\mathbb Z}\, \backslash {\mathbb Y}$. For Fourier coefficients of functions $v\in L^2_{\Lambda (T)}({\mathbb R};{\mathbb C})$, we use the notation $v_Y=T^{-1}\int_0^Tv(t)\, e^{-iYt}\, dt$, $Y\in \omega {\mathbb Z}$. We write
$$
{\mathfrak L}({\mathbb Y}^{\prime })=\bigl\{ v\in L^{2,\, even}_{\Lambda (T)}({\mathbb R};{\mathbb C}): \, v_Y=0\ \, for \ \, Y\in {\mathbb Y}\bigr\} .
$$
Denote by $\widehat P_{{\mathbb Y}^{\prime }}$ the orthogonal projection in the space $(L^{2,\, even}_{\Lambda (T)}({\mathbb R};{\mathbb C}),\| \cdot \| _{L^2_{\Lambda (T)}})$ onto the subspace ${\mathfrak L}({\mathbb Y}^{\prime })$;
$\widehat P_{\mathbb Y}\doteq \widehat I-\widehat P_{{\mathbb Y}^{\prime }}$ where $\widehat I$ is the identity operator on $L^{2,\, even}_{\Lambda (T)}({\mathbb R};{\mathbb C})$. Let 
$$
{\omega }_*(t)=2 \cos \omega t,\quad t\in {\mathbb R}.
$$
\vskip 0.3cm

{\bf Lemma 11.} {\it For any functions ${\mathcal F}_j\in {\mathfrak L}({\mathbb Y}^{\prime })$, $j=1, \dots , m$, there exist unique functions $G_j\in {\mathfrak L}({\mathbb Y}^{\prime })\cap H^2_{\Lambda (T)}({\mathbb R};{\mathbb C})\subset {\mathfrak L}({\mathbb Y}^{\prime })\cap C_{\Lambda (T)}({\mathbb R};{\mathbb C})$ such that
$$
-\, \frac {d^2G_j}{dt^2}-B_0\sum\limits_{\mu \, =\, 1}^mC^{(m)}_{j\mu }G_{\mu }=B_0{\mathcal F}_j,\quad j=1,\dots ,m.
$$
Moreover, $G_j=\sum\limits_{\mu }\widehat S_{j\mu }{\mathcal F}_{\mu }$ where $\widehat S_{j\mu }: ({\mathfrak L}({\mathbb Y}^{\prime }),\| \cdot \| _{L^2_{\Lambda (T)}})\to ({\mathfrak L}({\mathbb Y}^{\prime })\cap C_{\Lambda (T)}({\mathbb R};{\mathbb C}),\| \cdot \| _{C_{\Lambda (T)}})$ are linear bounded operators. For all ${\mathcal F}_j\in 
{\mathfrak L}({\mathbb Y}^{\prime })\cap C_{\Lambda (T)}({\mathbb R};{\mathbb C})$,
$$
\sum\limits_j\| G_j\| _{C_{\Lambda (T)}}\leqslant CT^{-1/2}\sum\limits_j\| {\mathcal F}_j\| _{L^2_{\Lambda (T)}}\leqslant C\sum\limits_j\| {\mathcal F}_j\| _{C_{\Lambda (T)}}
$$
where $C=C(m)>0$. If ${\mathcal F}_j\in {\mathfrak L}({\mathbb Y}^{\prime })\cap L^2_{\Lambda (T)}({\mathbb R};{\mathbb R})$, $j=1,\dots ,m$, then $G_j\in {\mathfrak L}({\mathbb Y}^{\prime })\cap C_{\Lambda (T)}({\mathbb R};{\mathbb R})$. If ${\mathcal F}_j\in {\mathfrak L}({\mathbb Y}^{\prime })\cap C^{\infty }_{\Lambda (T)}({\mathbb R};{\mathbb C})$, $j=1,\dots ,m$, then $G_j\in {\mathfrak L}({\mathbb Y}^{\prime })\cap C^{\infty }_{\Lambda (T)}({\mathbb R};{\mathbb R})$.}
\vskip 0.3cm

{\it Proof}. Let $n\in {\mathbb Z}\backslash \{ -1,1\} $ and let $Y=n\omega \in 2\pi \Lambda ^*(T)$. By the choice of the eigenvalue ${\lambda }^{(m)}$, the number $n^2{\omega }^2$ is not an eigenvalue of the matrix $B_0\widehat C^{(m)}$. Hence there are numbers $b_j^{(Y)}\in {\mathbb C}$, $j=1,\dots ,m$, such that $b_j^{(-Y)}=b_j^{(Y)}$,
$$
n^2{\omega }^2b_j^{(Y)}-B_0\sum\limits_{\mu }C^{(m)}_{j\mu }b_{\mu }^{(Y)}=B_0({\mathcal F}_j)_Y, \eqno(4.1)
$$
and
$$
\sum\limits_j|b_j^{(Y)}|\, \leqslant \, C(Y)\sum\limits_j|({\mathcal F}_j)_Y| \eqno(4.2)
$$
where $C(Y)>0$. This implies that, for all $Y=n\omega \in {\mathbb Y}^{\prime }$,
$$
\sum\limits_j(1+n^2)|b_j^{(Y)}|\, \leqslant \, C^{\prime }\sum\limits_j|({\mathcal F}_j)_Y|
$$
where $C^{\prime }=C^{\prime }(m)>0$. Then, for functions
$$
G_j\doteq \sum\limits_{Y\, \in \, {\mathbb Y}^{\prime }}b_j^{(Y)}\, e^{\, iYt}\, \in \, 
{\mathfrak L}({\mathbb Y}^{\prime })\cap C_{\Lambda (T)}({\mathbb R};{\mathbb C})
$$
we find
$$
\sum\limits_j\| G_j\| _{C_{\Lambda (T)}}\, \leqslant \, \sum\limits_j\sum\limits_{Y\, \in \, {\mathbb Y}^{\prime }}|(G_j)_Y| \eqno(4.3)
$$ $$
\leqslant \, \sum\limits_j \, \biggl( \, \sum\limits_{Y\, \in \, 2\pi \Lambda ^*(T)}(1+n^2)^{-2}\biggr) ^{1/2}\biggl( \,  \sum\limits_{Y\, \in \, {\mathbb Y}^{\prime }}(1+n^2)^2|(G_j)_Y|^2\biggr) ^{1/2}
$$ $$
\leqslant \, mC^{\prime }\biggl( \, \sum\limits_{Y\, \in \, 2\pi \Lambda ^*(T)}(1+n^2)^{-2}\biggr) ^{1/2}\biggl( \,  \sum\limits_{Y\, \in \, {\mathbb Y}^{\prime }}
\biggl( \, \sum\limits_j|({\mathcal F}_j)_Y|^2\biggr) \biggr) ^{1/2}
$$ $$
\leqslant \, C\sum\limits_j \, \biggl( \,  \sum\limits_{Y\, \in \, {\mathbb Y}^{\prime }}|({\mathcal F}_j)_Y|^2\biggr) ^{1/2}=\, CT^{-1/2}\sum\limits_j\| {\mathcal F}_j\| _{L^2_{\Lambda (T)}}\leqslant \, \sum\limits_j\| {\mathcal F}_j\| _{C_{\Lambda (T)}}
$$
where $C=mC^{\prime }\biggl( \, \sum\limits_{Y\, \in \, 2\pi \Lambda ^*(T)}(1+n^2)^{-2}\biggr) ^{1/2}$. The functions $G_j$ satisfy all conditions of Lemma 11. \hfill $\square $
\vskip 0.3cm

We use the notation $U_r=\{ z\in {\mathbb C}:|z|<r\} $, $\overline U_r=\{ z\in {\mathbb C}:|z|\leqslant r\} $, $r>0$.
\vskip 0.3cm

{\bf Lemma 12.} {\it If, for some $r>0$, functions
$$
U_r\ni z\mapsto {\mathcal F}_j(z)\in ({\mathfrak L}({\mathbb Y}^{\prime }),\| \cdot \| _{L^2_{\Lambda (T)}}),\quad j=1,\dots ,m,
$$
are analytic, then the functions
$$
U_r\ni z\mapsto G_j(z)=\sum\limits_{\mu \, =\, 1}^m\widehat S_{j\mu }{\mathcal F}_{\mu }(z)\in
({\mathfrak L}({\mathbb Y}^{\prime })\cap C_{\Lambda (T)}({\mathbb R};{\mathbb C}),\| \cdot \| _{C_{\Lambda (T)}}) \eqno(4.4)
$$
also are analytic.}
\vskip 0.3cm

{\it Proof}. It follows from (4.1) and (4.2) that Fourier coefficients of the functions $G_j(z)$ are analytic functions. On the other hand, for all $R\geqslant 0$ and $z\in U_r$, analogously to inequality (4.3), we obtain
$$
\sum\limits_j\, \sum\limits_{|Y|\, >\, R}|(G_j(z))_Y| \,
\leqslant \, mC^{\prime }\biggl( \, \sum\limits_{n\, \in \, {\mathbb Z}:\, |n|\omega \, >\, R}(1+n^2)^{-2}\biggr) ^{1/2}T^{-1/2}\sum\limits_j\| {\mathcal F}_j\| _{L^2_{\Lambda (T)}}. 
$$
Therefore, for all $r_1\in (0,r)$, the partial sums
$$
\sum\limits_{Y\, \in \, 2\pi \Lambda ^*(T):\, |Y|\, \leqslant \, R}(G_j(z))_Y\, e^{\, iYt}
$$
converge in th space $({\mathfrak L}({\mathbb Y}^{\prime })\cap C_{\Lambda (T)}({\mathbb R};{\mathbb C}),\| \cdot \| _{C_{\Lambda (T)}})$ to the functions $G_j(z)$ as $R\to +\infty $ uniformly in $z\in \overline U_{r_1}$ and consequently functions (4.4) are analytic. \hfill $\square $ 
\vskip 0.3cm

Let $B_0>0$. We choose a number $r>0$. Some bounds from above for the number $r$ will be given in what follows. For all $\varepsilon \in U_r$, let us represent complex solutions of system of equations (0.11) with $\tau =\tau (\varepsilon )\in {\mathbb C}$ in the form
$$
v_j=v_j(\varepsilon ;\cdot )=\varepsilon a_jw_*+{\varepsilon }^2w_j+{\varepsilon }^3b_jw_*, \quad j=1,\dots ,m, \eqno(4.5)
$$
where $a=(a_1,\dots ,a_m)^T\in {\mathbb R}^m$ is the eigevector of the matrix $\widehat C^{(m)}$ with the eigenvalue ${\lambda }^{(m)}$, $\| a\| _{{\mathbb R}^m}=1$, $a_1>0$, $b=b(\varepsilon )=(b_1,\dots ,b_m)^T\in {\mathbb C}^m$ and $(b,a)_{{\mathbb C}^m}=0$, $w_j=w_j(\varepsilon ;\cdot )\in {\mathfrak L}({\mathbb Y}^{\prime })\cap C^{\infty }_{\Lambda (T)}({\mathbb R};{\mathbb C})$. The last summand in (4.5) is absent if $m=1$. 
The (complex) functions ${\mathfrak w}_j(\varepsilon ;\cdot )$ in (0.10) are chosen as ${\mathfrak w}_j(\varepsilon ;\cdot )=w_j(\varepsilon ;\cdot )+\varepsilon b_jw_*(\cdot )$, $j=1,\dots ,m$. 

Since $a$ is the eigenvector of the matrix  $B_0\widehat C^{(m)}$ with the eigenvalue $\omega ^2$ and $\widehat P_{\, {\mathbb Y}}w^2_*$ is the zero function, system of equations (0.11) is equivalent to systems 

$$
\biggl( \omega ^2b_j-B_0\sum\limits_{\mu }C^{(m)}_{j\mu }b_{\mu }-\tau \omega ^2a_j\biggr) w_* \eqno(4.6)
$$ $$
=\, B_0\widehat P_{\, {\mathbb Y}}\biggl( \, \sum\limits_{\mu }C^{(m)}_{j\mu }a_{\mu }w_*w_{\mu }\biggr) +\frac {B_0}6\, \widehat P_{\, {\mathbb Y}}\biggl( \, \sum\limits_{\mu }C^{(m)}_{j\mu }a^3_{\mu }\biggr) w^3_*+\varepsilon B_0\widehat P_{\, {\mathbb Y}}{\mathcal F}_j^{(0)},\quad j=1,\dots ,m,
$$ $$
-\frac {d^2w_j}{dt^2}\, -\, B_0\sum\limits_{\mu }C^{(m)}_{j\mu }w_{\mu }=\frac {B_0}2\, \biggl( \, \sum\limits_{\mu }C^{(m)}_{j\mu }a^2_{\mu }\biggr) w^2_*+\varepsilon B_0 
\widehat P_{\, {\mathbb Y}^{\prime }}{\mathcal F}_j^{(1)}, \quad j=1,\dots ,m. \eqno(4.7)
$$
which are derived from (0.11) under the orthogonal projections $\widehat P_{\, {\mathbb Y}}$ and $\widehat P_{\, {\mathbb Y}^{\prime }}$. The functions ${\mathcal F}_j^{(\sigma )}$, $\sigma =0,1$, can be explicitly found from (0.11) by expansion of the function ${\mathbb C}\ni \zeta \mapsto e^{\, \zeta }$ in power series. This functions satisfy following properties (1), (2), and (3).

(1) For all $\varepsilon ,\tau ,b_{\nu }\in {\mathbb C}$ and $w_{\nu }\in {\mathfrak L}({\mathbb Y}^{\prime })\cap C_{\Lambda (T)}({\mathbb R};{\mathbb C})$, $\nu =1,\dots ,m$,
the functions ${\mathcal F}_j^{(\sigma )}={\mathcal F}_j^{(\sigma )}(\varepsilon ,\tau ,\{ b_{\nu }\} ,\{ w_{\nu }\} )$, $\sigma =0,1$,  belong to the space $C^{\, even }_{\Lambda (T)}({\mathbb R};{\mathbb C})$. If $w_{\nu }\in {\mathfrak L}({\mathbb Y}^{\prime })\cap C^{\infty }_{\Lambda (T)}({\mathbb R};{\mathbb C})$, $\nu =1,\dots ,m$, then ${\mathcal F}_j^{(\sigma )}\in C^{\infty ,\, even}_{\Lambda (T)}({\mathbb R};{\mathbb C})$.

(2) If the numbers $\tau ,b_{\nu }$ and the functions $w_{\nu }\in {\mathfrak L}({\mathbb Y}^{\prime })\cap C_{\Lambda (T)}({\mathbb R};{\mathbb C})$, $\nu =1,\dots ,m$, depend on a number $\varepsilon \in {\mathbb C}$, and functions ${\mathbb C}\ni \varepsilon \mapsto \tau \in {\mathbb C}$, ${\mathbb C}\ni \varepsilon \mapsto b_{\nu }\in {\mathbb C}$, and ${\mathbb C}\ni \varepsilon \mapsto w_{\nu }\in \bigl( {\mathfrak L}({\mathbb Y}^{\prime })\, \cap \, C_{\Lambda (T)}({\mathbb R};{\mathbb C}),\| \cdot \| _{C_{\Lambda (T)}}\bigr) $ are analytic, then functions ${\mathbb C}\ni \varepsilon \mapsto {\mathcal F}_j^{(\sigma )}(\varepsilon ,\tau ,\{ b_{\nu }\} ,\{ w_{\nu }\} )\in \bigl( C^{\, even}_{\Lambda (T)}({\mathbb R};{\mathbb C}),\| \cdot \| _{C_{\Lambda (T)}}\bigr) $ also are analytic.

(3) For any $R>0$ there exist numbers $C_R^{(\sigma )}>0$, $\sigma =0,1$, such that for all numbers $\varepsilon ,\tau ^{\prime },\tau ^{\prime \prime },b_{\nu }^{\prime },b_{\nu }^{\prime \prime }\in \overline U_R$ and all functions $w_{\nu }^{\prime },w_{\nu }^{\prime \prime }\in {\mathfrak L}({\mathbb Y}^{\prime })\cap C_{\Lambda }({\mathbb R};{\mathbb C})$, for which $\| w_{\nu }^{\prime }\| _{C_{\Lambda (T)}}\leqslant R$ and $\| w_{\nu }^{\prime \prime }\| _{C_{\Lambda (T)}}\leqslant R$, $\nu =1,\dots ,m$, inequalities
$$
\sum\limits_j\| {\mathcal F}_j^{(\sigma )}(\varepsilon ,\tau ^{\prime },\{ b_{\nu }^{\prime }\} ,\{ w_{\nu }^{\prime }\} )-{\mathcal F}_j^{(\sigma )}(\varepsilon ,\tau ^{\prime \prime },\{ b_{\nu }^{\prime \prime }\} ,\{ w_{\nu }^{\prime \prime }\} )\| _{C_{\Lambda (T)}}
$$ $$ 
\leqslant \, C_R^{(\sigma )}\biggl( |\tau ^{\prime }-\tau ^{\prime \prime }|+\sum\limits_{\nu }|b_{\nu } ^{\prime }-b_{\nu }^{\prime \prime }|+\sum\limits_{\nu }\|w_{\nu }^{\prime }-w_{\nu }^{\prime \prime }\| _{C_{\Lambda (T)}}\biggr) \eqno(4.8)
$$
hold.

Lemma 11 implies that system of equations (4.7) is equivalent to the system
$$
w_j={\mathbb W}_j+\varepsilon {\mathcal F}_j^{(2)},\quad j=1,\dots ,m, \eqno(4.9)
$$
where
$$
{\mathbb W}_j\doteq \frac 12\, \sum\limits_{\mu }\, \biggl( \, \sum\limits_{\nu }C^{(m)}_{\mu \nu }a^2_{\nu }\biggr) \widehat S_{j\mu }w^2_*,\quad {\mathcal F}^{(2)}_j\doteq \sum\limits_{\mu }\widehat S_{j\mu }\widehat P_{\, {\mathbb Y}^{\prime }}{\mathcal F}^{(1)}_{\mu }.
$$
Therefore, if system of equations (4.7) is valid, then system of equations (4.6) is equivalent to the system
$$
\biggl( \omega ^2b_j-B_0\sum\limits_{\mu }C^{(m)}_{j\mu }b_{\mu }-\tau \omega ^2a_j\biggr) w_*=B_0\widehat P_{\, {\mathbb Y}}\widetilde {\mathbb W}_j+\varepsilon B_0\widehat P_{\, {\mathbb Y}}{\mathcal F}_j^{(3)},\quad j=1,\dots ,m, \eqno(4.10)
$$
where
$$
\widetilde {\mathbb W}_j\doteq \sum\limits_{\mu }C^{(m)}_{j\mu }a_{\mu }w_*{\mathbb W}_{\mu }+\frac 16\, \biggl( \, \sum\limits_{\mu }C^{(m)}_{j\mu }a^3_{\mu }\biggr) w^3_*,\quad {\mathcal F}_j^{(3)}\doteq \sum\limits_{\mu }C^{(m)}_{j\mu }a_{\mu }w_*{\mathcal F}_{\mu }^{(2)}+{\mathcal F}^{(0)}_j.
$$
The functions ${\mathcal F}_j^{(\sigma )}={\mathcal F}_j^{(\sigma )}(\varepsilon ,\tau ,\{ b_{\nu }\} ,\{ w_{\nu }\} )$, $\sigma =2,3$, as well 
as the functions ${\mathcal F}_j^{(\sigma )}={\mathcal F}_j^{(\sigma )}(\varepsilon ,\tau ,\{ b_{\nu }\} ,\{ w_{\nu }\} )$, $\sigma =0,1$, satisfy properties (1), (2), and (3)
(for the functions ${\mathcal F}_j^{(2)}$, property (2) is fulfilled by Lemma 12). 

Since $\| w_*\| ^2_{L^2_{\Lambda (T)}}=2T$,
it follows from Lemma 10 that equations (4.10) hold for $\varepsilon \in {\mathbb C}$ iff
$$
\tau =\tau ^{(0)} +\frac {\varepsilon }{2T}\, \sum\limits_{\mu }\tau _{\mu }(w_*,{\mathcal F}^{(3)}_{\mu })_{L^2_{\Lambda (T)}}, \eqno(4.11)
$$ $$
b_{\nu }=b_{\nu }^{(0)}+\frac {\varepsilon }{2T}\, \sum\limits_{\mu }R_{\nu \mu }(w_*,{\mathcal F}^{(3)}_{\mu })_{L^2_{\Lambda (T)}},
\quad \nu =1,\dots ,m, \eqno(4.12)
$$
where
$$
\tau ^{(0)}\doteq \frac 1{2T}\, \sum\limits_{\mu }\tau _{\mu }(w_*,\widetilde {\mathbb W}_{\mu })_{L^2_{\Lambda (T)}},\quad b_{\nu }^{(0)}\doteq \frac 1{2T} \, \sum\limits_{\mu }R_{\nu \mu }(w_*,\widetilde {\mathbb W}_{\mu })_{L^2_{\Lambda (T)}}.
$$

Denote $w^{(0)}_{\nu }\doteq {\mathbb W}_{\nu }$, $ \nu =1,\dots ,m$, and sequentially for $j=1,2,\dots $ let us define numbers $\tau ^{(j)}=\tau ^{(j)}(\varepsilon )$, $b_{\nu }^{(j)}=b_{\nu }^{(j)}(\varepsilon )$ and functions $w_{\nu }^{(j)}=w_{\nu }^{(j)}(\varepsilon ;\cdot )$. If they are already found for some $j\in {\mathbb Z}_+$, then we set
$$
\tau ^{(j+1)}=\tau ^{(0)} +\frac {\varepsilon }{2T} \, \sum\limits_{\mu }\tau _{\mu }\big( w_*,{\mathcal F}^{(3)}_{\mu }(\varepsilon ,\tau ^{(j)},\{ b_s^{(j)}\} ,\{ w^{(j)}_s\} )\bigr) _{L^2_{\Lambda (T)}},
$$ $$
b_{\nu }^{(j+1)}=b_{\nu }^{(0)} +\frac {\varepsilon }{2T}\, \sum\limits_{\mu }R_{\nu \mu }\big( w_*,{\mathcal F}^{(3)}_{\mu }(\varepsilon ,\tau ^{(j)},\{ b_s^{(j)}\} ,\{ w^{(j)}_s\} )\bigr) _{L^2_{\Lambda (T)}},
$$ $$
w_{\nu }^{(j+1)}=w_{\nu }^{(0)}+\varepsilon {\mathcal F}^{(2)}_{\nu }(\varepsilon ,\tau ^{(j)},\{ b_s^{(j)}\} ,\{ w^{(j)}_s\} ).
$$
Then $w_{\nu }^{(j)}(\varepsilon ;\cdot )\in {\mathfrak L}({\mathbb Y}^{\prime })\cap C^{\infty }_{\Lambda (T)}({\mathbb R};{\mathbb C})$ for all $\varepsilon \in {\mathbb C}$, and functions ${\mathbb C}\ni \varepsilon \mapsto \tau ^{(j)}(\varepsilon )$, ${\mathbb C}\ni \varepsilon \mapsto b_{\nu }^{(j)}(\varepsilon )$, ${\mathbb C}\ni \varepsilon \mapsto w_{\nu }^{(j)}(\varepsilon ;\cdot )\in \bigl( {\mathfrak L}({\mathbb Y}^{\prime })\cap C_{\Lambda (T)}({\mathbb R};{\mathbb C}),\| \cdot \| _{C_{\Lambda (T)}}\bigr) $ are analytic  (and $\sum\limits_{\nu }\overline b^{(j)}_{\nu }(\varepsilon )a_{\nu }=0$), $\nu =1,\dots ,m$, $j\in {\mathbb N}$. If $\varepsilon \in {\mathbb R}$, then $\tau ^{(j)}(\varepsilon ),b_{\nu }^{(j)}(\varepsilon )\in {\mathbb R}$ and $w^{(j)}_{\nu }(\varepsilon ;\cdot )\in C^{\infty }_{\Lambda (T)}({\mathbb R};{\mathbb R})$.

Inequality (3.1) and inequality (4.8) (which holds for the functions ${\mathcal F}^{(2)}_j$ and ${\mathcal F}^{(3)}_j$) imply that there is a number $r>0$ such that for all $\varepsilon \in \overline U_r$ and all $j\in {\mathbb N}$
$$
\bigl| \tau ^{(j+1)}(\varepsilon )-\tau ^{(j)}(\varepsilon )\bigr| + \sum\limits_{\nu }\, \bigl| b_{\nu }^{(j+1)}(\varepsilon )-b_{\nu }^{(j)}(\varepsilon )\bigr| + \sum\limits_{\nu }\, \bigl\| w_{\nu }^{(j+1)}(\varepsilon ;\cdot )-w_{\nu }^{(j)}(\varepsilon :\cdot )\bigr\| _{C_{\Lambda (T)}}
$$ $$
\leqslant \, \frac 12\, \biggl( \bigl| \tau ^{(j)}(\varepsilon )-\tau ^{(j-1)}(\varepsilon )\bigr| + \sum\limits_{\nu }\, \bigl| b_{\nu }^{(j)}(\varepsilon )-b_{\nu }^{(j-1)}(\varepsilon )\bigr| + \sum\limits_{\nu }\, \bigl\| w_{\nu }^{(j)}(\varepsilon ;\cdot )-w_{\nu }^{(j-1)}(\varepsilon :\cdot )\bigr\| _{C_{\Lambda (T)}}\biggr) .
$$
Hence for any $\varepsilon \in \overline U_r$, there exist numbers $\tau (\varepsilon )$, $b_{\nu }(\varepsilon )\in {\mathbb C}$ and functions $w_{\nu }(\varepsilon ;\cdot )\in {\mathfrak L}({\mathbb Y}^{\prime })\, \cap \, C^{\infty }_{\Lambda (T)}({\mathbb R};{\mathbb C})$ such that $\tau ^{(j)}(\varepsilon )\to \tau (\varepsilon )$, $b_{\nu }^{(j)}(\varepsilon )\to b_{\nu }(\varepsilon )$, and $\| w_{\nu }^{(j)}(\varepsilon ;\cdot )-w_{\nu }(\varepsilon;\cdot )\| _{C_{\Lambda (T)}}\to 0$ as $j\to +\infty $. Because the convergence is uniform in $\varepsilon \in \overline U_r$, it follows that the functions $U_r\ni \varepsilon \mapsto \tau (\varepsilon )$, $U_r\ni \varepsilon \mapsto b_{\nu }(\varepsilon )$, and $U_r\ni \varepsilon \mapsto w_{\nu }(\varepsilon ;\cdot )\in \bigl( {\mathfrak L}({\mathbb Y}^{\prime })\, \cap \, C^{\infty }_{\Lambda (T)}({\mathbb R};{\mathbb C}),\| \cdot \| _{C_{\Lambda (T)}}\bigr) $ are analytic. For all $\varepsilon \in \overline U_r$, the numbers $\tau (\varepsilon)$, $b_j(\varepsilon )$ and the functions $w_j(\varepsilon ;\cdot )$ satisfy equations (4.11), (4.12), and (4.9), therefore  $w_j(\varepsilon ;\cdot )\in H^2_{\Lambda (T)}({\mathbb R};{\mathbb C})\subset C_{\Lambda (T)}({\mathbb R};{\mathbb C})$ and functions (4.5) are solutions of system (1.11). In addition, the number $r>0$ can be chosen sufficiently small such that the inequality ${\varepsilon }^2|\tau (\varepsilon )|<1$ holds for all $\varepsilon \in \overline U_r$. For real numbers $\varepsilon \in (-r,r)$, the numbers $\tau (\varepsilon )$, $b_j(\varepsilon )$ also are real and the functions $u_j(\varepsilon ;\cdot )$ belong to $H^2_{\Lambda (T)}({\mathbb R};{\mathbb R})$. Since under the condition $u_j(\varepsilon ;\cdot )\in H^{2n}_{\Lambda (T)}({\mathbb R};{\mathbb R})$, $n\in {\mathbb N}$, $j=1,\dots ,m$, it follows from (1.11) that $u_j(\varepsilon ;\cdot )\in H^{2(n+1)}_{\Lambda (T)}({\mathbb R};{\mathbb R})$, we have
$$
u_j(\varepsilon ;\cdot )\in \bigcap\limits_{n\, \in \, {\mathbb N}}H^{2n}_{\Lambda (T)}({\mathbb R};{\mathbb R})=C^{\infty }_{\Lambda (T)}({\mathbb R};{\mathbb R}).
$$
The proof of Theorem 7 is complete.

\end{document}